# An automatic multi-tissue human fetal brain segmentation benchmark using the Fetal Tissue Annotation Dataset

Kelly Payette[1,2], Priscille de Dumast[3,10], Hamza Kebiri[3,10], Ivan Ezhov[4], Johannes C. Paetzold[4], Suprosanna Shit[4], Asim Iqbal[2,5], Romesa Khan[2,6], Raimund Kottke[7], Patrice Grehten[7], Hui Ji[1], Levente Lanczi[8], Marianna Nagy[8], Monika Beresova[8], Thi Dao Nguyen[9], Giancarlo Natalucci[9,11], Theofanis Karayannis[5], Bjoern Menze[4], Meritxell Bach Cuadra[3,10,12], Andras Jakab[1,2]

1 Center for MR Research, University Children's Hospital Zurich; 2 Neuroscience Center Zurich, University of Zurich/ETH Zurich; 3 Department of Diagnostic and Interventional Radiology, Lausanne University Hospital and University of Lausanne, Lausanne, Switzerland; 4 Image-Based Biomedical Imaging Group, Technical University of Munich; 5 Brain Research Institute, University of Zurich; 6 Institute for Biomedical Engineering, UZH/ETH Zurich; 7 Diagnostic Imaging, University Children's Hospital Zurich; 8 University of Debrecen, Hungary; 9 Newborn Research Zurich, Department of Neonatology, University Hospital and University of Zurich, Zurich, Switzerland; 10 Medical Image Laboratory Analysis (MIAL), Centre d'Imagerie BioMédicale (CIBM), University of Lausanne, Lausanne, Switzerland; 11 Larsson-Rosenquist center for neurodevelopment, growth and nutrition of the newborn, University Hospital and University of Zurich, Zurich, Switzerland; 12 Signal Processing Laboratory (LTS5), Ecole Polytechnique Fédérale de Lausanne (EPFL)

**Corresponding author:** Kelly Payette, Email: kelly.payette@kispi.uzh.ch

## Abstract

It is critical to quantitatively analyse the developing human fetal brain in order to fully understand neurodevelopment in both normal fetuses and those with congenital disorders. To facilitate this analysis, automatic multi-tissue fetal brain segmentation algorithms are needed, which in turn requires open databases of segmented fetal brains. Here we introduce a publicly available database of 50 manually segmented pathological and non-pathological fetal magnetic resonance brain volume reconstructions across a range of gestational ages (20 to 33 weeks) into 7 different tissue categories (external cerebrospinal fluid, grey matter, white matter, ventricles, cerebellum, deep grey matter, brainstem/spinal cord). In addition, we quantitatively evaluate the accuracy of several automatic multi-tissue segmentation algorithms of the developing human fetal brain. Four research groups participated, submitting a total of 10 algorithms, demonstrating the benefits the database for the development of automatic algorithms.

## Background & Summary

Congenital disorders are one of the leading causes of neonatal mortality worldwide (http://www.who.int/gho/child_health/mortality/causes/en/). In order to detect and treat congenital disorders, prenatal maternal and fetal healthcare is of the utmost importance. Fetal imaging with ultrasound has been an essential tool in prenatal care for many years. Recently, fetal magnetic resonance imaging (MRI) has started to emerge as an important supplemental tool for providing information about the developing fetus when the ultrasound image is unclear, or for more accurate diagnosis of congenital disorders[1–3]. Fetal MRI, especially of the brain, has been shown to be useful in the diagnosis of many congenital disorders such as spina bifida, intrauterine growth retardation, congenital heart disease, lissencephaly, and corpus callosum anomalies[3–7].

Fetal MRI is a challenging imaging modality requiring both clinical and technical expertise as the fetus is not sedated and is able to move freely. Ultra-fast MRI sequences such as T2-weighted single shot fast spin echo (ssFSE) acquire low resolution images very quickly and can be acquired in all planes. Super-resolution (SR) reconstruction algorithms can then be applied in order to combine several low-resolution images into a single high resolution volume, incorporating outlier rejection and motion correction strategies[8–14], which can then be used for further quantitative analysis.

Automated quantification of the highly complex and rapidly changing brain morphology in MRI data could improve the diagnostic and decision-making process. An initial step for the quantification of fetal brain morphology and volumetry is image segmentation. It is clinically relevant to analyse information such as shape or volume of the developing cortex, cerebellum, brainstem, white matter and cerebrospinal fluid spaces, as many congenital disorders cause subtle changes to these tissue compartments[6,15,16]. Existing growth data is mainly based on normally developing brains [17–19], and we lack growth data for many pathologies and congenital disorders. The automatic segmentation of the developing human brain would be a first step in being able to perform such an analysis, as manual segmentation is both time-consuming and prone to human error. However, segmenting these SR fetal brain volumes is still challenging due to rapid changing anatomy of the fetal brain, remaining motion or blurry artefacts and partial volume effects.

Atlas-based methods of segmenting brain tissues have been developed[20,21], however these methods require having an atlas, which currently only exist for normally developing fetuses. In order to study pathological brains, especially in the cases of severe pathology, alternate atlases would be required. Recently, convolutional neural networks have also been used to segment the fetal brain into different tissue types, but only for low-resolution scans in the coronal direction[22], which has limited use. Other work has shown it to be possible to segment a single class within high-resolution fetal brain volumes, but multi-class semantic segmentation is not addressed[23]. The major bottleneck of developing segmentation algorithms for medical imaging is the lack of data, either availability of atlases for atlas-based segmentation, or training data for supervised machine learning methods. The field of fetal MRI has so far been understudied due to the lack of accessible, curated and annotated ground truth data.

In this paper, we present a unique dataset (FeTA – Fetal Tissue Annotation and Segmentation Dataset), consisting of SR fetal brain volumes that have been reconstructed from multiple clinically acquired ssFSE scans, as well as a manual segmentation of each brain into seven different brain tissues. The dataset consists of both normal and pathological brains with an age range of 20 to 33 gestational weeks. This dataset is meant to serve as a training dataset for automatic multi-class semantic or instance segmentation algorithms, and subsets of the dataset have already been utilized in research. One study used the grey matter segmentations of all cases to train a network with topological loss, and the other used all labels of the non-pathological cases to investigate the role of transfer learning in fetal brain segmentation[24,25]. We first outline the methods used to acquire and process the data, as well as how the labels were created. Next, we present several algorithms from four different research groups which used the dataset to create automatic segmentation methods based on multi-atlas segmentation methods and deep learning. In addition, we would like to announce a segmentation challenge using the FeTA dataset, inviting any interested groups to create automatic fetal brain segmentation algorithms for evaluation. It is important that we create accurate and reproducible methods of analysing the developing fetal brain in order to support decision-making and prenatal planning, and better understand the underlying causes of congenital disorders.

## Methods

**Image Acquisition:** Fetal MRI was acquired in 50 pregnant women at the University Children's Hospital Zurich between 2016 and 2019. For each subject, multiple MRI scans of the brain were acquired on either a 1.5T or 3T clinical GE whole-body scanner (Signa Discovery MR450 and MR750) either using an 8-channel cardiac coil or body coil. T2-weighted SSFSE sequences were acquired with an in-plane resolution of 0.5mmx0.5mm and a slice thickness of 3 to 5mm. The sequence parameters were the following: TR: 2000-3500 ms, TE: 120 ms (minimum), flip angle: 90°, sampling percentage 55%. Field of view (200-240 mm) and image matrix (1.5T: 256x224; 3T: 320x224) were adjusted depending on the gestational age (GA) and size of the fetus. Imaging plane was oriented relative to the fetal brain and axial, coronal and sagittal images were acquired.

**Subject characteristics:** The images were either acquired as part of the clinical diagnostic routine, or as a part of an ongoing research study. The subjects were chosen to be part of the dataset as they cover both pathological and non-pathological fetal brains at a variety of GAs. The average GA in weeks of the subjects at the time of MRI was 26.9 weeks with a standard deviation of 3.3 weeks (calculated GA). The subjects can be split into two categories: non-pathological brains (18 subjects), and pathological brains with spina bifida either before or after fetal spinal lesion repair surgery (32 subjects). See Table 1 for a summary of the diagnostic categories and GA of fetuses. All study participants gave informed written consent for their MRI datasets to be used for research purposes in an anonymized form. Mothers of the healthy fetuses participating in the BrainDNIU study were prospectively informed about the study by members of the research team and gave written consent for their participation. Mothers of all other fetuses included in the current work were scanned as part of their routine clinical care and gave informed written consent for the re-use of their data for research purposes. The ethical committee of the Canton of Zurich, Switzerland approved the prospective and retrospective studies that collected and analysed the MRI data (Decision numbers: 2017-00885, 2016-01019, 2017-00167), and a waiver for an ethical approval was acquired for the release of an anonymous dataset for non-medical research purposes.

**Image Processing and Quality Assessment:** For each subject, we manually reviewed the acquired fetal brain images for quality in order to compile a stack of images. Each stack consisted of at least one brain scan in each orientation, with more scans included when available. The number of scans in each stack ranged between 3 and 13. Every image in the stack was then reoriented to a standard plane and a mask was created of the fetal brain using a semi-automated atlas-based custom MeVisLab (MeVis Medical Solutions AG, Bremen, Germany) module[9,26]. An SR reconstruction algorithm was then applied to each subject's stack of images and brain masks, creating a 3D SR volume of brain morphology[9,27] with an isotropic resolution of 0.5mm*0.5mm*0.5mm.

In addition, the SR reconstructions are of varying quality in order to mimic potential real-world cases, as fetal movement is uncontrolled and random. The quality of the reconstructions was judged by 3 independent reviewers with experience in fetal MRI (RKo, PG and MBC) using a Likert scale from 1 to 3, where a rating of 1 was considered to be poor quality, with motion and blurring, rendering it unusable for segmentation or diagnostic purposes, 2 is overall good quality with some remaining blurring, but usable for further analysis, and a rating of 3 was considered to be excellent. The correlation of the reviewers was calculated using the Gwet AC coefficient using R (v4.0.2)[28], and the percent agreement was found to be 0.89[29,30]. As the ratings are ordinal data, the median of the ratings is considered to be the final rating of the SR volume. See Figure 1 for an overview of the quality of the SR reconstructions, and Table 2 for the quality rating for each reconstructed SR volume.

*Table 1: Subject Characteristics split into Training and Testing, as well as Pathological and Non-Pathological fetal SR brains volumes.*

| Gestational Age* | Training (Image + labels) | | Testing (Image only) | | Total |
|---|---|---|---|---|---|
| | Non-pathological | Pathological | Non-pathological | Pathological | |
| 20 | 0 | 1 | 0 | 0 | 1 |
| 21 | 0 | 0 | 1 | 1 | 2 |
| 22 | 1 | 2 | 0 | 1 | 4 |
| 23 | 1 | 4 | 0 | 0 | 5 |
| 24 | 1 | 2 | 0 | 0 | 3 |
| 25 | 0 | 3 | 0 | 1 | 4 |
| 26 | 2 | 3 | 0 | 0 | 5 |
| 27 | 2 | 5 | 0 | 2 | 9 |
| 28 | 1 | 2 | 1 | 1 | 5 |
| 29 | 1 | 2 | 0 | 0 | 3 |
| 30 | 0 | 1 | 0 | 1 | 2 |
| 31 | 2 | 0 | 0 | 0 | 2 |
| 32 | 3 | 0 | 1 | 0 | 4 |
| 33 | 1 | 0 | 0 | 0 | 1 |
| Total: | 15 | 25 | 3 | 7 | 50 |

* Gestational age is given in weeks and is the postmenstrual age, measured from the first day of the last normal menstrual period. The given gestational age refers to a full week range (e.g. 20 + 0/7 – 20 + 6/7).

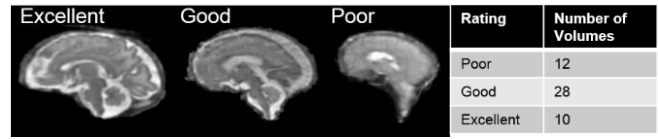

*Figure 1: Left: Examples of fetal brain SR volumes with different quality ratings; Right: Quality ratings overview (1: poor quality; 2: good quality; 3: excellent quality)*

*Table 2: Median quality rating and number of scans used to create each SR volume for each subject*

| Subject Number | Median Quality Rating | Number of Scans used to create SR volume | Subject Number | Median Quality Rating | Number of Scans used to create SR volume |
|---|---|---|---|---|---|
| 01 | 2 | 5 | 26 | 2 | 8 |
| 02 | 1 | 5 | 27 | 3 | 12 |
| 03 | 2 | 6 | 28 | 3 | 8 |
| 04 | 2 | 6 | 29 | 2 | 4 |
| 05 | 1 | 8 | 30 | 2 | 13 |
| 06 | 3 | 3 | 31 | 2 | 9 |
| 07 | 1 | 10 | 32 | 2 | 11 |
| 08 | 2 | 4 | 33 | 3 | 9 |
| 09 | 1 | 3 | 34 | 3 | 8 |
| 10 | 3 | 7 | 35 | 2 | 7 |
| 11 | 2 | 5 | 36 | 2 | 6 |
| 12 | 2 | 5 | 37 | 2 | 4 |
| 13 | 3 | 5 | 38 | 2 | 5 |
| 14 | 2 | 4 | 39 | 3 | 7 |
| 15 | 2 | 4 | 40 | 2 | 5 |
| 16 | 2 | 4 | 41 | 2 | 7 |
| 17 | 1 | 5 | 42 | 1 | 5 |
| 18 | 2 | 9 | 43 | 3 | 4 |
| 19 | 2 | 6 | 44 | 2 | 8 |
| 20 | 1 | 6 | 45 | 1 | 7 |
| 21 | 1 | 9 | 46 | 1 | 11 |
| 22 | 1 | 5 | 47 | 3 | 6 |
| 23 | 1 | 9 | 48 | 2 | 7 |
| 24 | 2 | 8 | 49 | 2 | 6 |
| 25 | 2 | 6 | 50 | 2 | 12 |

**Manual Segmentation:** Manual segmentation was performed on the reconstructed volume for each subject using the Draw tool in 3D Slicer[31,32]. The labelling scheme and anatomical definitions used are based on those used for the brain tissue segmentation within the Developing Human Connectome Project pipeline for neonatal datasets[33]. The following seven labels were used: external cerebrospinal fluid (CSF), grey matter (GM), white matter (WM), ventricles, cerebellum, deep grey matter, and brainstem/spinal cord. Each label was created separately, with the annotator segmenting every second to every third slice for each label in the axial plane, except for the cerebellum and the brainstem/spinal cord, which were segmented in the sagittal plane. Due to the large annotation workload, different people annotated certain labels for every case. KP annotated the external CSF space; MN annotated the gray matter and brainstem; MB annotated the white matter and cerebellum; LL annotated the deep gray matter; AJ annotated the ventricles and performed the final corrections. The final label map was created by post-processing these sparse annotations to create a single fully segmented fetal brain for each subject. For interpolating the sparsely annotated label maps, we used the Python implementation of the ITK nD Morphological Contour Interpolation algorithm, enforcing interpolation along the plane the given structure was annotated[34]. After post-processing, each fetal brain was reviewed by an expert and small corrections were made either on the original annotations or the reconstructed full fetal brain annotation. We provide a detailed definition of the annotated anatomical structures in each label, a description of the manual annotation process and the post-processing of the labels in the Supplementary Information. An example of manual segmentations can be seen in Figure 2.

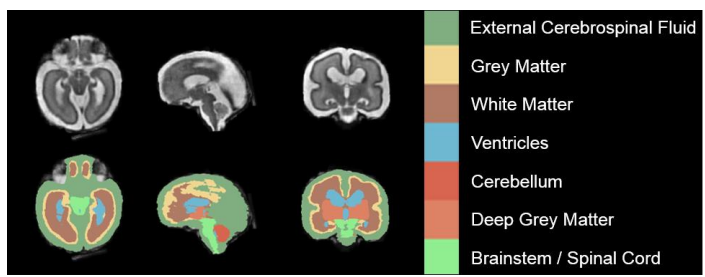

*Figure 2: Example of Manual Segmentation (Dark Green: external cerebrospinal fluid; yellow: GM; brown: WM; blue: ventricles; bright red: cerebellum; light red: deep GM: bright green: brainstem/spinal cord)*

**Inter-Rater Variability:** Nine volumes were manually segmented three times by three different annotators with experience in fetal MRI (the additional two segmentations were performed by KP and HJ annotating all labels). The same method and the same software (3D Slicer [31,32]) for segmentation was used. The segmentations underwent the same post-processing steps, all as outlined in the Supplementary Information. Annotator 1 was compared to Annotator 2, Annotator 2 was compared to Annotator 3, and Annotator 1 was compared to Annotator 3, resulting in 3 different sets of metrics, which were then averaged together. All metrics were generated using the same methodology as was used in comparing the generated label maps to the 'ground truth'.

No consensus delineation was used in the manual segmentations in the FeTA dataset, it is based on a single segmentation. This is due to the large time and resource requirement for performing multiple segmentations for all cases within the dataset.

We wanted to begin to understand the differences between annotators segmenting the fetal brains. It has been shown that volume overlap variability of eight raters performing segmentation tasks on a variety of lesions and organs is between 8-21% and depends on structures and pathologies. They also conclude that two to three observers is not sufficient to establish the full range of inter-observer variability (for any two observers the variability was found to be 5-57%)[35]. Therefore we acknowledge that with our three annotations of nine volumes, we are unable to make concrete conclusions regarding the variability. Instead we wish to emphasize the existence of the problem of high variability between annotators, as well as the 'noisiness' of the labels themselves, as sometimes there is no clear correct label. In addition, the label definitions themselves may depend on the training and experience of the annotator. In the future, if more alternate segmentations are created, methods such as curve registration could be used to estimate a 'true label' based on the multiple annotations[36].

We looked at the difference in annotators for each label class across all images, as well as split into the categories 'Non Pathological cases' (n=3) and 'Pathological cases' (n=6), as well as 'Excellent Quality SR volumes' (n=3), 'Good Quality SR volumes' (n=3) and 'Poor Quality SR volumes' (n=3).

There was variability across the three annotators for the nine volumes which had multiple segmentations, as seen in Figure 3. It is challenging to say whether or not this variability is acceptable, as there are no other public datasets available containing multiple fetal brain segmentations of the same volume with which to compare. And as mentioned above, the sample size is small. However, the results are as we expected. Tissue classes where the boundaries are poorly defined had a high amount of variability, such as the external CSF space and GM in the low quality images (external CSF DSC: 0.13±0.11; GM DSC: 0.45±0.10). High quality volumes (external CSF DSC: 0.88±0.02; GM DSC: 0.71±0.03) and non-pathological volumes (external CSF DSC: 0.88±0.02; GM DSC: 0.72±0.04) had less variability, with DSC of 0.88±0.02 in the external CSF space for both groups, and DSC of 0.71±0.03 in the GM of the high quality volumes and 0.72±0.04 in the non-pathological volumes (there was only one volume that was part of both groups). In light of this high variability, consensus delineations of the fetal brain should be investigated for future releases of the dataset with an increased number of different observer annotations.

## Data Records

The reconstructions and manual segmentations used within this paper can be found on Synapse under Fetal Tissue Annotation Challenge FeTA[37] Dataset (website: http://dx.doi.org/10.7303/syn23747212), and is organized in the Brain Imaging Dataset Structure (BIDS) format. The released database consists of 40 SR reconstructions and their corresponding manual segmentation. The manual segmentation consists of the combined label map containing all tissue types, as well as the individual tissue labels. In addition, 10 SR reconstructions are included without the manual segmentation for validation purposes. The database also contains the guidelines used by the annotators as well as the code that was used for the creation of the final manual segmentations.

## Technical Validation

**Segmentation methods:** Here we present several automatic multi-tissue segmentation algorithms for SR volumes of the fetal brain that have used this dataset to train/develop the algorithm. Four research groups participated, submitting a total of 10 algorithms, demonstrating the benefits the database for the development of automatic algorithms. Each algorithm was evaluated using evaluation criteria outlined below.

<u>Multi-Atlas Segmentation Method</u> (Priscille de Dumast, Hamza Kebiri, Meritxell Bach Cuadra):

Atlas-based segmentation techniques rely on image registration processes, which seek to find the voxel to voxel alignment of two images[38]. One of the images is an atlas from which a label map of the structures of interest is available, and the other one is the target subject's image to be segmented. Once the spatial transformation between the grey-level images is found, the target label map (segmentation) is estimated with the propagation of the atlas labels into the target space. In multi-atlas segmentation (MAS), the above-mentioned method is repeated for a set of atlases[39]. Afterwards, label fusion strategies allow for the combination of the propagated label maps, eventually attributing more weights – globally or locally – to one or another selected atlas. In the context of fetal brain MRI, both atlas based and multi-atlas based segmentation methods are used[20,40,41].

In our approach, all forty training subjects – hence with manual annotations – are initially considered as atlases, although a more specific criteria-based selection of atlas candidates is performed. Each of the atlas' super-resolution (SR) image is classified into one of the three categories: bad (remaining motion, blurred, unusable for segmentation purposes), acceptable (overall good quality with some blurring but usable for segmentation purposes) or excellent. SR reconstructions estimated as bad were considered not good enough for registration purposes and thus discarded. In fetal brain segmentation, the selection of appropriate atlases is challenging due to the major morphological development of the fetal brain that is happening along the course of gestation. Intuitively, atlas candidates that are selected are the ones more likely to present similar morphology than the target subject. Therefore, in our approach only those within two weeks younger or older were registered to the subject's space. Finally, the limitation on the number of available atlases did not allow for pathology-based discrimination criteria, so both normal developing and spina bifida subjects were used as atlases (see Figure 4 for an overview of atlases used in label fusion).

To proceed from global alignment of MR volumes to local deformation, a three-level registration is performed, using the Advanced Normalization Tools (ANTS)[42], in the following order, increasing the freedom of the transformation: rigid, affine and finally non-linear symmetric diffeomorphic. We used

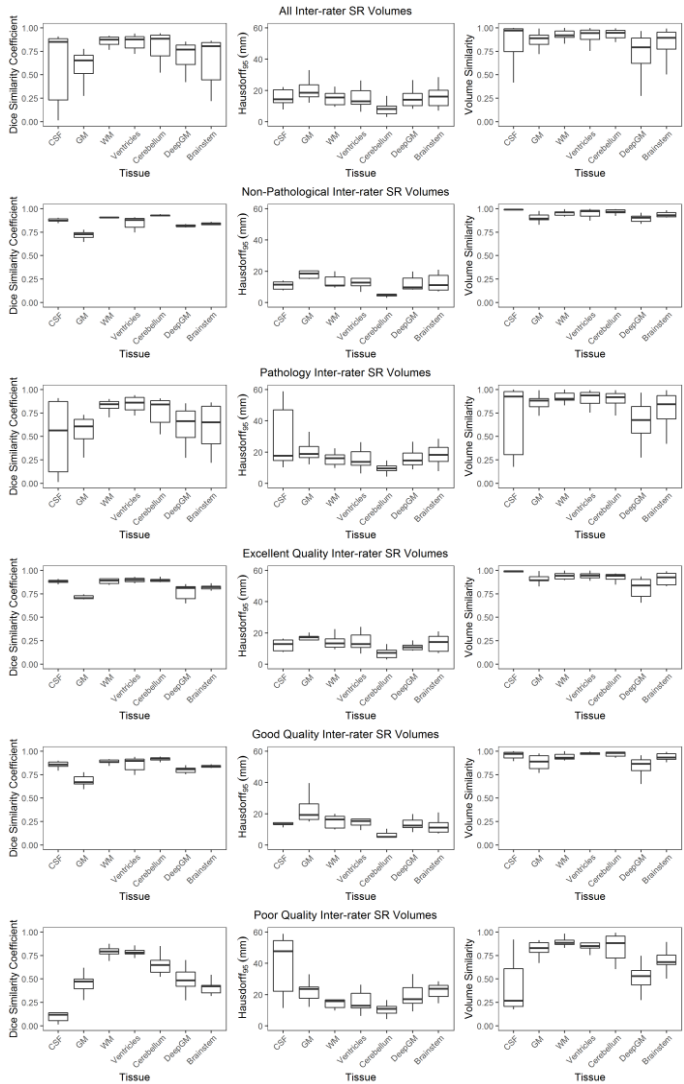

*Figure 3: Analysis of the 3 annotator segmentations of all 9 volumes averaged together, and split into the categories of normal SR volumes, pathological SR volumes, excellent quality SR volumes, good quality SR volumes, and poor quality SR volumes*

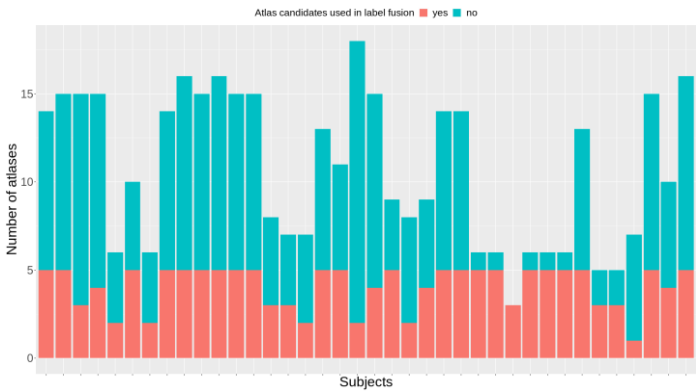

*Figure 4: Number of atlas candidates (up to 2 GA difference and acceptable SR quality) vs number of atlases (5 best ranked NCC and NCC ≥ 0.8) used in the label fusion step, for the training dataset, using a leave-one-out approach*

the intensity-based normalized cross-correlation (NCC) as similarity registration metric. Consequently, we will define registration as being sufficiently good for being part of the label fusion when its NCC is equal to or higher than 0.8 and the atlases that do not reach this NCC value are discarded. We use the NiftySeg implementation for NCC computation[43]. We limit the maximum number of atlases used for the label fusion to 5 (having the highest NCC). A local weighted voting algorithm is used for the final label fusion step[41] in which weighting is based on the NCC computed on a Gaussian kernel of size 10 voxels.

KispiU (Kelly Payette, Andras Jakab):

Three 2D U-Nets[44] were trained separately in each orientation (axial, sagittal, coronal), creating three separate networks. The images being fed into the network were 256x256, 1 channel, and a batch size of 16 was used. An Adam optimizer, a learning rate of 10E-5, L2 regularization, ReLu activation, and batch normalization after each convolutional layer were used. In addition, a dropout layer after each block of convolutional layers was used. The network was programmed in Keras and trained on an Nvidia Quadro P6000. Cross-validation was used with 80% of the images used for training and 20% used for testing, shuffling after each epoch. The network was trained for 200 epochs with early stopping, and the weights from the best performing epoch were kept. The generalized Dice coefficient was used as a loss function[45]. Data augmentation (flipping, 0-360° rotation, adding Gaussian noise) was also utilized.

A 3D U-Net with the same architecture as the 2D U-Net was also trained. The images were fed into the network in patches of 64x64x64 voxels. The same loss functions, optimizers, number of epochs, learning rates, data augmentation, and epochs were used. A batch size of 8 was used due to memory constraints. 3D network was trained on order to compare the results of a 2D network to a 3D network. This method is a collection of 4 neural networks: a 2D U-Net for each of the axial, sagittal, and coronal directions, as well as a 3D U-Net, in order to investigate whether a 3D or 2D network is better for this task. Each of the four networks was individually trained and evaluated.

SeBRe (Asim Iqbal, Romesa Khan, Theofanis Karayannis):

The FeTA dataset contains 40 brain stacks in 3D with each brain of size 256×256×256 - covering the sagittal, coronal and axial planes. Along with the brain images, each stack also contains manual annotations of 7 different brain regions (256×256×256). In order to make the dataset compatible for SeBRe[46], we first converted the 3D brain stacks into 2D brain images (256×256) covering the top-down axial plane. Furthermore, we converted the manual annotations into binary mask (256×256) images to develop the training and validation set for Mask RCNN-based SeBRe architecture. To measure the performance of SeBRe on FeTA dataset, we used 80% of the brains for training (randomly selected) and remaining 20% for validation. The final classification threshold for predicting brain region masks was set to 0.5 confidence to avoid weak predictions. To compile the results on the 10 unseen validation brains, a binary threshold was set on the predicted segmentations (label maps) before compiling the quantitative scores. The code generated in segmenting brain regions will be available online at https://github.com/itsasimiqbal/SeBRe.

SeBRe is designed by optimization of Mask R-CNN[47] deep neural network architecture, constructed using a convolutional backbone that comprises of the first five stages of the very deep ResNet101[48] and Feature Pyramid Network (FPN)[49] architectures. The network architecture is the same as described in the original study[46]. The input feature map is processed by a convolutional neural network through a Region Proposal Network (RPN) in a sliding-window fashion which forwards the predicted $n$ Regions of Interest (ROI) from each window to the Mask R-CNN 'heads' based on the Fully Pyramidal Network (FPN). FPN is multi-headed: 'classifier' head acts as an identifier, 'regressor' head, similar to [50] and [51] detects the brain region and computes region-specific bounding boxes, and 'segmenter' predicts the masks using Fully Convolutional Network (FCN)[52]. We apply transfer learning i.e. training on the FeTA brain section dataset is initialized with pre-trained weights for the Microsoft COCO dataset and we freeze all the layers during training except the heads of the network.

The classifier output layer returns a discrete probability distribution $[n, N]$, here $N$ brain regions are predicted along with a background class. The regressor output layer gives the four (x-coordinate, y-coordinate, width, height) bounding-box regression offsets to be applied for each class. The network is trained using a stochastic gradient descent algorithm that minimizes a multi-task loss ($L$) corresponding to each labelled ROI:

$$L = L_{cls} + L_{reg} + L_{mask}$$

where $L_{cls}$, $L_{reg}$ and $L_{mask}$ are the region classification, bounding box regression and mask prediction losses, respectively, as defined below.

$$L(p_i, q_i) = \frac{1}{n_{cls}} \sum_i L_{cls}(p_i, p_i^*) + \mu \frac{1}{n_{reg}} \sum_i L_{reg}(q_i, q_i^*)$$

where $p_i$ is the probability of the $i^{th}$ proposed ROI, or anchor, enclosing an object. $p_i^*$ denotes if the anchor is positive ($p_i^* = 1$) or negative ($p_i^* = 0$). Vector $q_i$ represents the four coordinates, characterizing the predicted anchor bounding box, whereas vector $q_i^*$ represents the coordinates for the ground-truth box corresponding to a positive anchor. $L_{cls}$ for each anchor is calculated as log loss for two class labels (brain region vs. no brain region). $L_{reg}$ is a regression loss function robust to the outliers, $n_{cls}$ and $n_{reg}$ are the normalization parameters for classification and regression losses, respectively, weighted by a parameter $\mu$[50]. $L_{mask}$ is computed as average cross-entropy loss for per-pixel binary classification, applied to each ROI[47].

IBBM (Ivan Ezhov, Johannes C. Paetzold, Suprosanna Shit, Bjoern Menze):
We use three 2D U-Net[44,53–55] segmentation architectures with some modifications. We implemented the 2D segmentation architecture separately for slices of all three orientations (axial, coronal and sagittal) of the 3D dataset (analogous to Guha Roy et al[56]). Our encoder is made of ResNet34 backbone pretrained on the ImageNet. As a loss for all three networks we used a sum of Dice and binary cross entropy. We used data augmentations such as flipping, rotation, scaling, and shifting. All networks are implemented in Pytorch using the Adam optimizer. Network were trained for 200 epochs, keeping the weights configuration that performs best on the validations set.

As discussed in the previous paragraph, we train, validate and test on the three orientations. Thereby, we generate three predictions per pixel. As a merging strategy we employ a weighted voting per pixel, see Equation below. Here P denotes the per pixel probabilities and $\alpha$, $\beta$ and $\gamma$ are weights for the orientations:

$$P_{final} = \alpha \times P_{axial} + \beta \times P_{coronal} + \gamma \times P_{saggital}$$

The weights $\alpha$, $\beta$ and $\gamma$ are tunable parameters. For our segmentation we carried out a grid on these weights in steps of 0.1 for all $\alpha + \beta + \gamma = 1$. We carried out the grid search on a validation set of 10 3D volumes and found optimal weights of $\alpha = 0.4$, $\beta = 0.4$ and $\gamma = 0.2$. We submitted each of the three individual 2D U-Nets as well as the weighted combination in order to validate the weightings.

**Method Evaluation Criteria:** Forty of the fifty volumes and corresponding manual segmentations were made available to the four different groups for use as training data/atlas creation for the creation of an automatic segmentation method. The remaining 10 volumes were given to each group, excluding the label map. These 10 volumes were used as a validation set in order to evaluate each algorithm. Each group submitted their generated label maps of the 10 validation volumes for evaluation. Evaluation criteria were designed in order to best evaluate automatic segmentation methods while avoiding common challenge pitfalls as outlined in [57]. As different metrics quantify various aspects of the quality of the segmentation, three different categories of metrics were used: an overlap-based metric, a volumetric metric, and a distance metric. Each metric was calculated for each label class within every volume. All evaluations will be performed using the EvaluateSegmentation Tool[58]. The following metrics were used for evaluation:

Dice Similarity Coefficient: The Dice similarity coefficient (DSC) was used to look at the overlap between the segmentations[59]. This metric is the most commonly-used metric to evaluate the quality of segmentations in medical imaging. It directly compares the automatic (new) segmentation and the ground truth segmentation, and is not impacted by outliers in the data[58]. It is defined as

$$DSC = \frac{2\,|GT \cap NS|}{|GT| + |NS|}$$

where GT is defined as the ground truth segmentation, and NS is defined as the new segmentation being evaluated.

Volume Similarity: Volumetric similarity (VS) is a metric that compares the volumes of the two segmentations. As several definitions exist for this metric, we take volumetric similarity as one minus the absolute volume difference between the ground truth and the new segmentation divided by the sum of the compared volumes:

$$VS = 1 - \frac{|GT_{vol} - NS_{vol}|}{GT_{vol} + NS_{vol}}$$

The VS is taken as a metric because volumetric quantification is a commonly used tool when looking at the development of the brain, and at differences between healthy and pathological brains. However, it cannot be looked at in isolation, as it is possible to have equal volumes without any overlap.

Hausdorff 95 Distance: The Hausdorff distance (HD) is a spatial metric and is also widely used for the evaluation of segmentations within medical imaging. It is especially helpful in evaluating the contours of segmentations as well as the spatial positions of the voxels. The HD between two finite point sets A and B is defined as

$$HD(A, B) = \max(h(A, B), h(B, A))$$

Where $h(A, B)$ is the directed HD, and is given by

$$h(A, B) = \max_{a \epsilon A} \min_{b \epsilon B} ||a - b||$$

The HD is not ideal in the presence of outliers. However, by using the 95th percentile of the HD (HD95), which uses the 95th quantile of distances instead of the maximum, thereby excluding possible outliers[58]. Therefore, HD95 (mm) was used.

**Rankings:** In order to evaluate and compare the ability of each algorithm to segment the fetal brain, a final score incorporating the three metrics (DSC, HD95, VS) was determined. All three metrics were calculated for each label within each of the corresponding predicted label maps of the fetal brain volumes in the testing set. The mean and standard deviation of each label were calculated, and the participating algorithms will be ranked from low to high (HD95), where the lowest score receives the highest scoring rank (best), and from high to low (DSC, VS), where the highest value will receive highest scoring rank (best). For each label, the three rankings were added together, and the algorithm with the highest ranking was ranked first. The ranked results can act as a benchmark for future multi-class segmentation algorithms.

In addition, the algorithms will be evaluated in the categories 'Non Pathological cases' and 'Pathological cases', as well as 'Excellent Quality SR volumes', 'Good Quality SR volumes' and 'Poor Quality SR volumes', with the identical ranking scheme for each category.

The overview of the metrics for each algorithm can be seen in Figure 5 with all labels combined. For the evaluation metrics separated by label number, see the Supplementary Information. The final rankings can be seen in Figure 5. The overview and rankings of the segmentations split by quality of the SR volumes can be found in Figure 6, and the overview and rankings of the segmentations separated into pathological and non-pathological SR volumes can be found in Figure 7.

There are some patterns evident in the results from all algorithms. The external CSF label was challenging to segment, mainly stemming from the younger pathological brains where the tissue boundaries are not clear, or is, in the case of some pathological cases, missing. The GM label also appeared to be challenging to properly segment. The cortical surface is constantly changing throughout gestation as a result of the ongoing neuronal migration and gyrification, resulting in changing contrast. Delineating the deep GM is also a challenge as the border between the deep GM and the surrounding structures is difficult to define on MRI, particularly in the region of the sub-thalamus and hypothalamus. The remaining structures had no consistent error patterns across all algorithms.

Multi-Atlas Segmentation Method: The multi-atlas segmentation method also performed very well, showing that neural networks may not always be superior. It seems that the multi-atlas method performed quite well when the quality of the SR was poor (see Figure 5). This is potentially because in a multi-atlas segmentation method, the atlas contains a highly probable shape of the structure already, so even if the structure is not clear in the image, the multi-atlas method is able to use its prior knowledge to provide a more accurate segmentation in non-ideal circumstances. This is also why the multi-atlas segmentation method scored high in the HD95 and VS metrics.

KispiU: Except in the poor-quality SR volumes, the Kispi 2D U-Net performed better than the 3D U-Net. While it was expected that the 3D U-Net would perform well as it contains 3D information, it comes at a cost of training size. If the training data is split into 2D slices, each image then contains 256 samples (as the volumes were 256x256x256). Whereas in the 3D U-Net, the image was split into 64x64x64 patches, resulting in 64 samples. To improve the 3D U-Net in the future, overlapping patches could be used.

SeBRe: Although the performance of SeBRe is competitive when compared to the other segmentation methods, its lower accuracy in predicting segmentations of the more complex brain regions could be explained by the neural network architecture. SeBRe is based on an instance segmentation-based network (i.e. Mask R-CNN) where the original feature detectors were designed to process higher-resolution images (e.g. MS-COCO dataset) than the SR fetal brain volumes on which SeBRe was evaluated here. Moreover, as SeBRe considers multiple appearances of similar objects as separate instances of the same object class, the ideal training regime would involve using separate masks for the spatially independent regions (e.g. ventricles) within the same object (brain tissue), an example of which is depicted for the case of fine-segmentation of sub-regions of a brain tissue (hippocampus)[46]. Adopting this strategy can be expected to enhance segmentation performance of the network on similar fetal brain segmentation tasks in the future.

IBBM: The combined method created by IBBM was the best ranking algorithm. Interesting to note is that while the axial and coronal directions have performed well, the same method trained in the sagittal direction did not perform as well. In all cases, the combined segmentations of IBBM outperformed the individual orientations.

**Final Evaluation/Discussion:** The dataset we are releasing is the first publicly released fetal brain dataset consisting of manually annotated volumes with multiple classes. However, there are limitations to the dataset. We chose to use SR volumes instead of the original radiological images. The SR method was arbitrarily chosen, and the differences between the existing SR methods have not been explored in detail which could have an impact on the success of a segmentation method as different reconstruction methods may result in SR volumes of differing quality. However, the usage of SR volumes instead of the original clinically acquired images has several benefits. It eliminates the need for the acquisition of 'in-plane' images required for basic biometric measurements, as the reconstruction results in a 3D volume. The integration of motion correction within the SR method allows for some movement to be present in the original images and is not present in the final SR volume. The main advantage of using 3D volumes instead of the original radiological images is that it allows for the eventual analysis of anatomy at a much finer scale, as the resolution of the volume is higher than that of the original scans. This results in a more complete picture of the developing human fetal brain.

The images included in the database have differing quality grades. The younger GAs, especially pathological cases, tend to be of a lower quality. While this is not ideal, this is often the case in clinical practice due to increased range of motion possible with smaller fetuses, and therefore it is seen as realistic. More pathologies could also be included in the database, which is a direction of future improvements to be investigated. In addition, only 9 volumes were annotated multiple times. This is a small number, and should be increased to better understand the inherent uncertainty within medical images, as well as the variance between annotators. If all SR volumes were segmented manually, this could be used to create a 'consensus segmentation', which has been used in other segmentation challenges[60], allowing the task to not be biased to any one particular annotator. In the cases of neural networks, another option would be for the networks to train on multiple annotations of the same volume, and then the network would learn the 'average' rater. In addition, priors could be developed from the results of this challenge and used in future segmentation algorithms.

The presented dataset is an important first step, and plans exist to update the dataset in order to address the limitations discussed. In the future, we plan to include fetal brain reconstructions performed with alternative SR methods, increased gestational ages, especially above 32 gestational weeks. There are also plans to increase the brain pathologies present in the database, as well as to increase the number of non-pathological brains. We hope to make this dataset as representative as possible of the clinical situation.

This clinical dataset is the first of its kind to be made public, and is shown here to be useful for developing automatic segmentation algorithms. A comprehensive knowledge of the developing human brain is a vital aspect of understanding how congenital disorders progress in utero. Through developing automatic segmentation methods, we aim to create tools to make this analysis a possibility in order to aid in prenatal planning for expecting parents.

## Usage Notes

Fetal Brain Segmentation Challenge: In order to foster further research on automatic fetal brain segmentation methods, we therefore are publicly releasing the dataset used within this analysis, and we invite other teams/groups to participate in the segmentation challenge (https://www.kispi.uzh.ch/fzk/de/abteilungen/uebersicht/mr-forschung/Seiten/fetal_developmental_imaging.aspx). Segmented methods must be fully automated with no other input than the fetal brain SR volume. Methods are free to evaluate the images in either 2D or 3D, but the evaluation will be performed in 3D. The entirety of the available training data does not necessarily need to be used, but when submitting final results, all testing samples must be submitted. If not all testing samples are submitted, those not submitted will result in a score of the worst metric value for each evaluation metric. This is done in order to avoid teams from artificially achieving high scores by only submitting segmentation results for 'easy' cases. The GA of the subject may be used as input.

Through the challenge and the release of the dataset, we aim to encourage groups within the medical image analysis community as well as the broader machine learning community to develop methods for the segmentation of the developing fetal human brain. Methods developed can be used to quantitatively analyse fetal brains in order to better understand their growth trajectory in both pathological and non-pathological cases. Through this analysis, we hope to better understand the underlying causes of congenital disorders, and aid parents in prenatal decision-making and planning.

## Code Availability

The code used during the development of the dataset can be found on Synapse under Fetal Tissue Annotation Challenge FeTA Dataset (website: http://dx.doi.org/10.7303/syn23747212)

<mark>Author's Original Version</mark> / manuscript under review

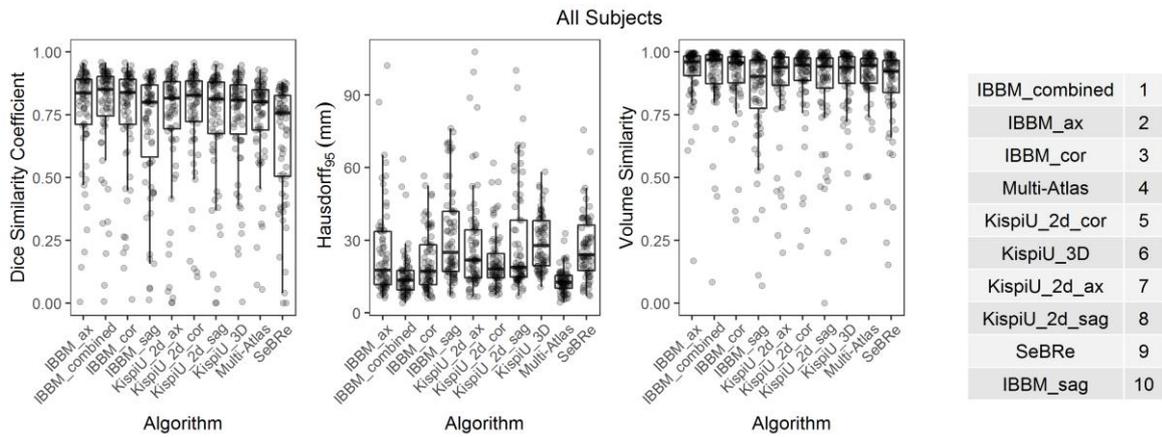

*Figure 5: Overview of the metrics (DSC, HD95, VS) of all subjects for each algorithm (all tissue labels combined), and the corresponding algorithm ranking.*

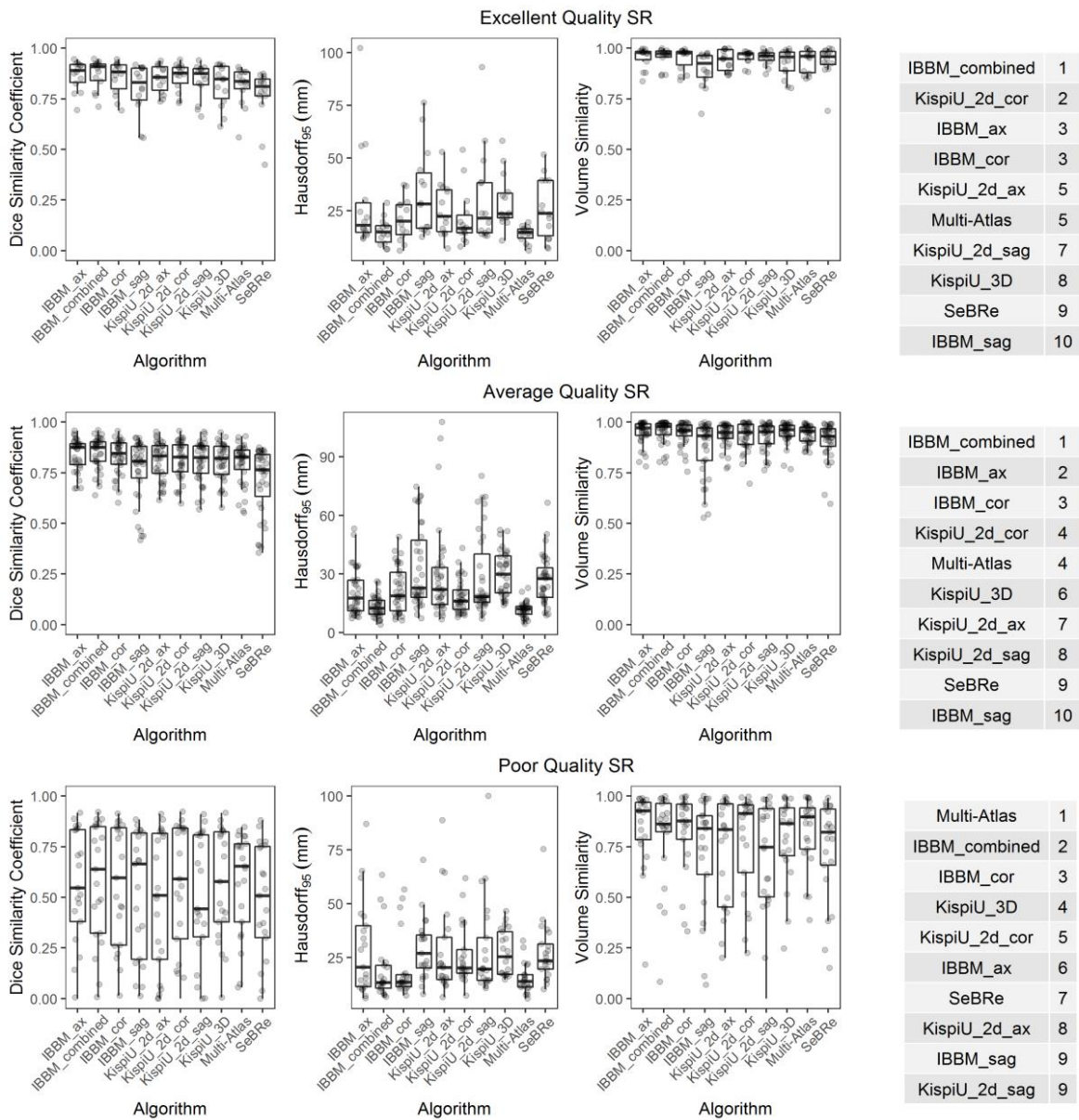

*Figure 6: An overview of the algorithms (all labels) evaluated on top: excellent quality SR volumes; middle: good quality SR volumes; and bottom: poor quality SR volumes, as well as the ranking for each of the quality levels. The poor-quality SR volumes have worse metrics (lower DSC and VS, higher HD95) than the average and excellent quality SR volumes, and the standard deviations are larger for every label. The difference in metrics between the average and excellent quality metrics is not noticeable.*

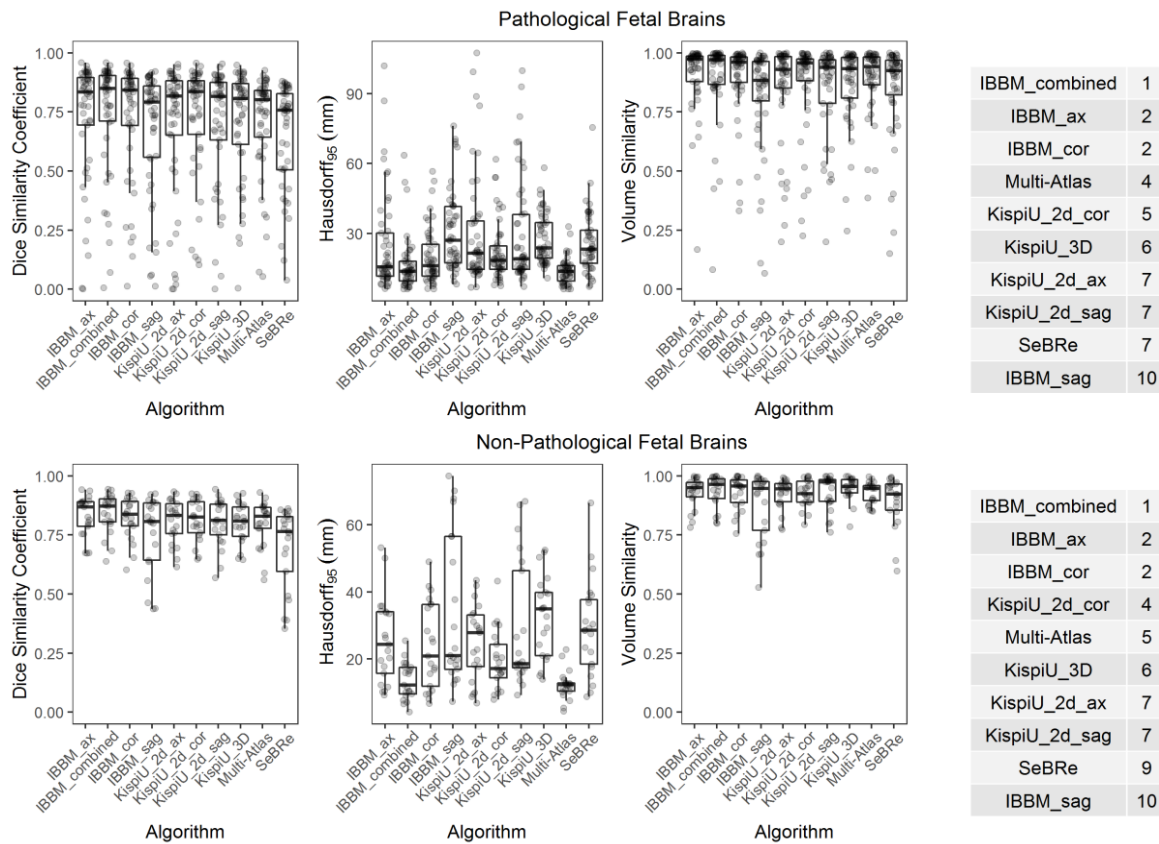

**Figure 7**: An overview of algorithms (all labels) evaluated on a) Pathological SR volumes and b) Non-pathological SR volumes, as well as the ranking for each. The segmentations of the normal SR volumes scored much higher than the pathological segmentations.

## Acknowledgements

The authors would like to acknowledge funding from the following funding sources: the OPO Foundation, the Prof. Dr. Max Cloetta Foundation, the Anna Müller Grocholski Foundation, the Foundation for Research in Science and the Humanities at the UZH, the EMDO Foundation, the Hasler Foundation, the FZK Grant, the Swiss National Science Foundation (project 205321-182602), and the ZNZ PhD Grant.

**This manuscript is the author's original version.**

# Supplementary Information

## Overview of each algorithm by label

In the following sections, the evaluation metrics for each algorithm separated by label number are displayed. (CSF: external cerebrospinal fluid; GM: Grey Matter; WM: White Matter)

**Multi-Atlas Segmentation**

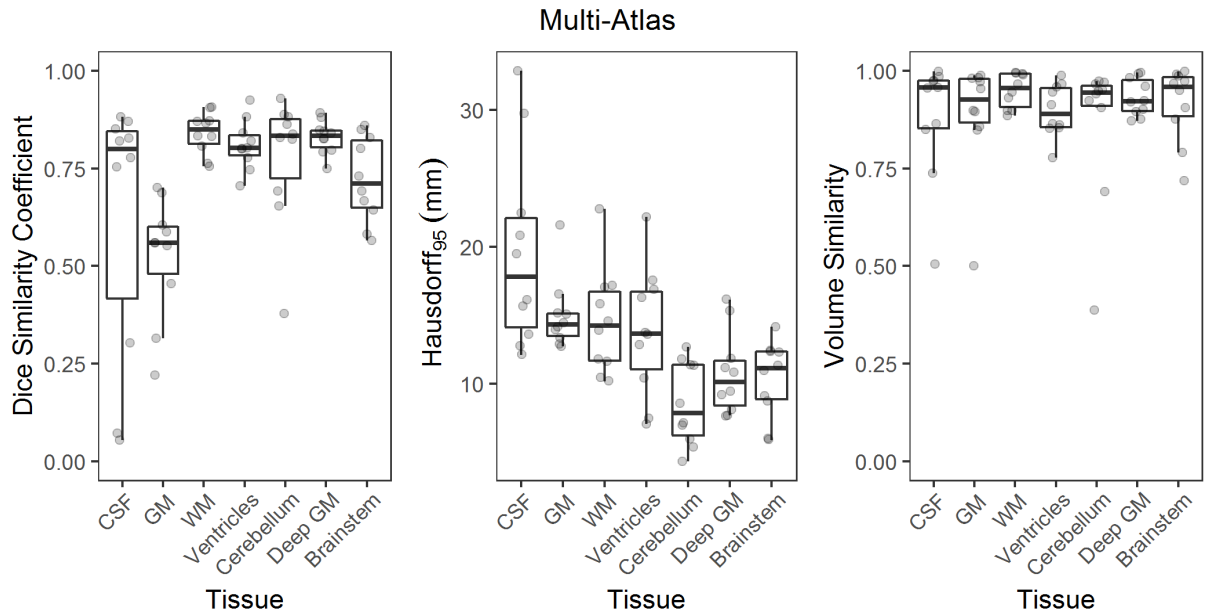

**SeBRe**

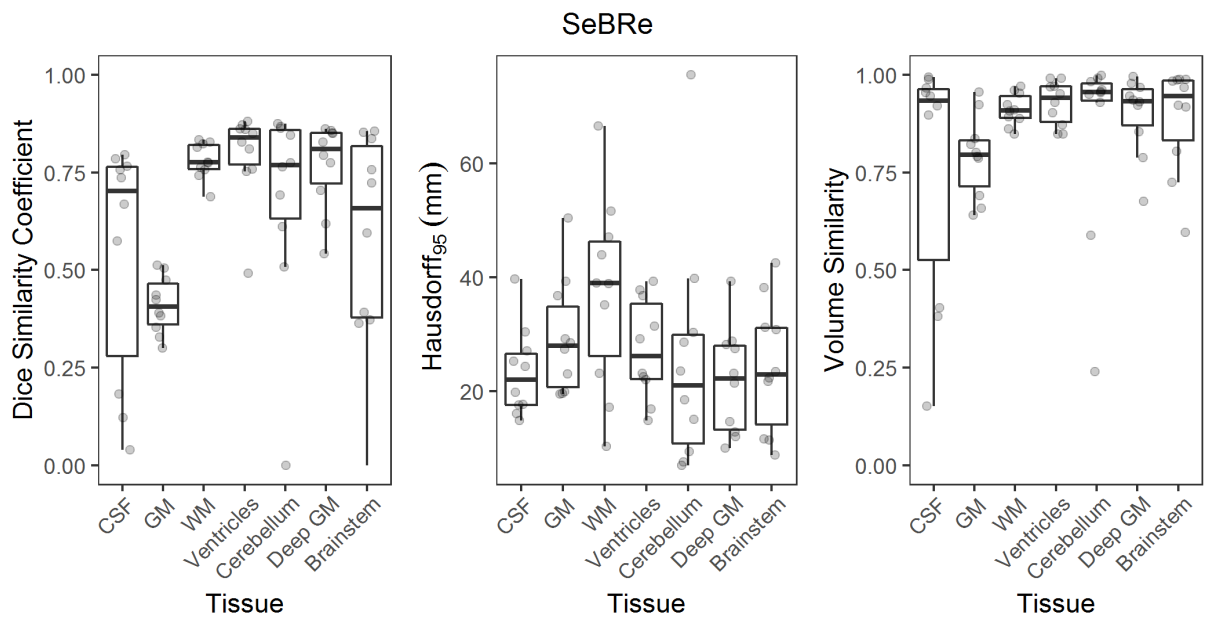

**IBBM Method – Axial Orientation**

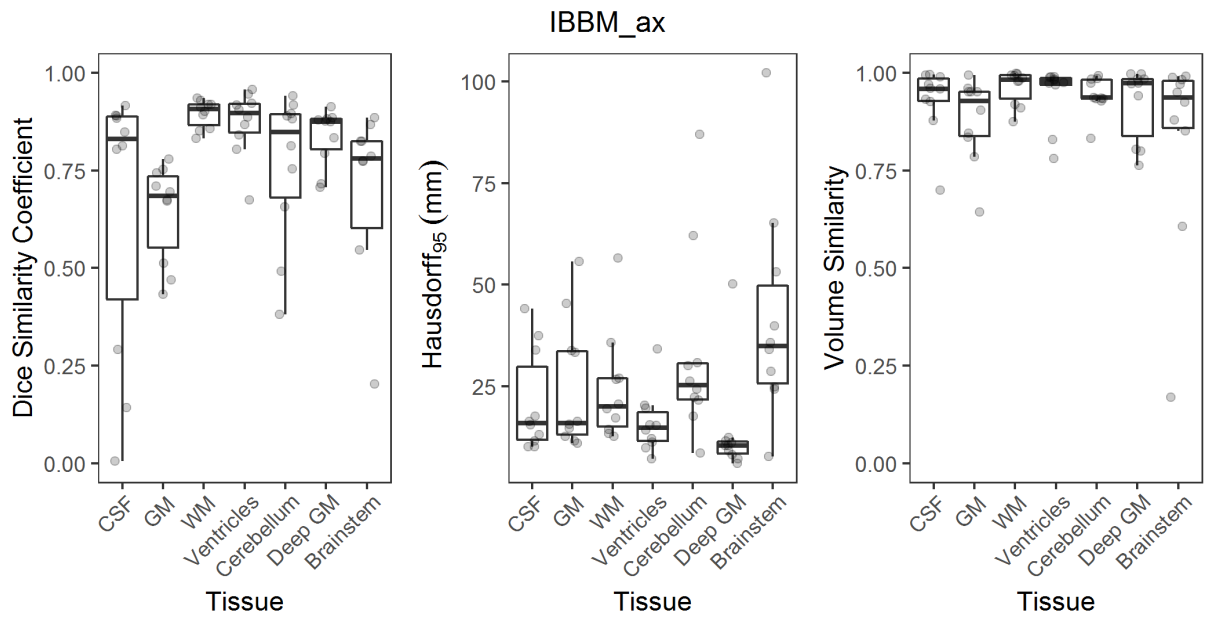

**IBBM Method – Sagittal Orientation**

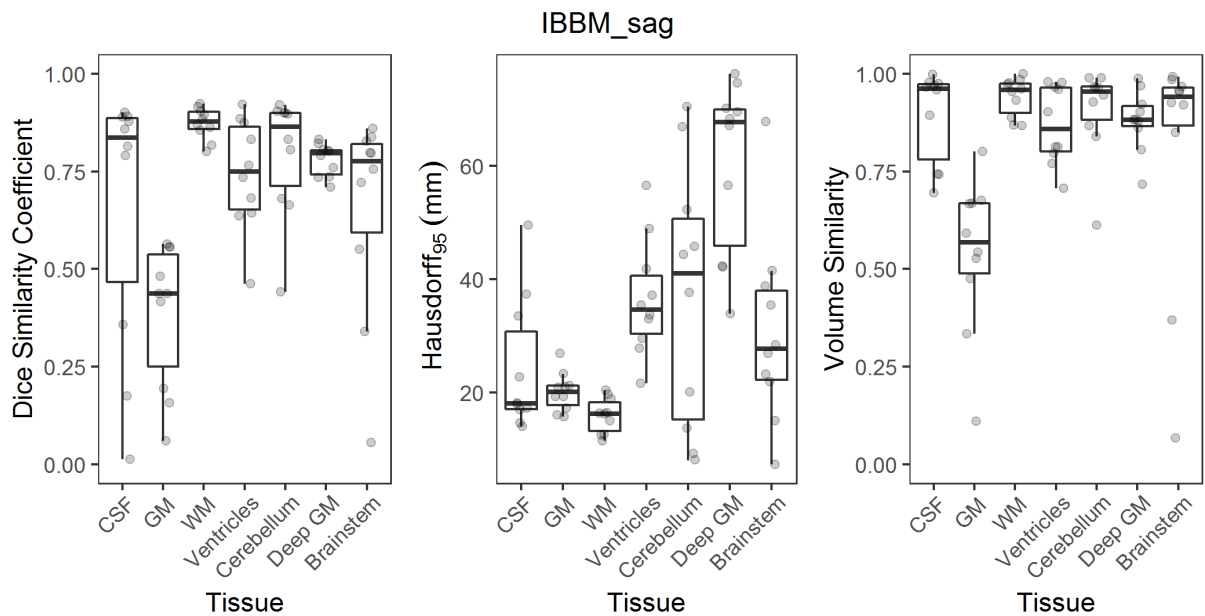

**IBBM Method – Coronal Orientation**

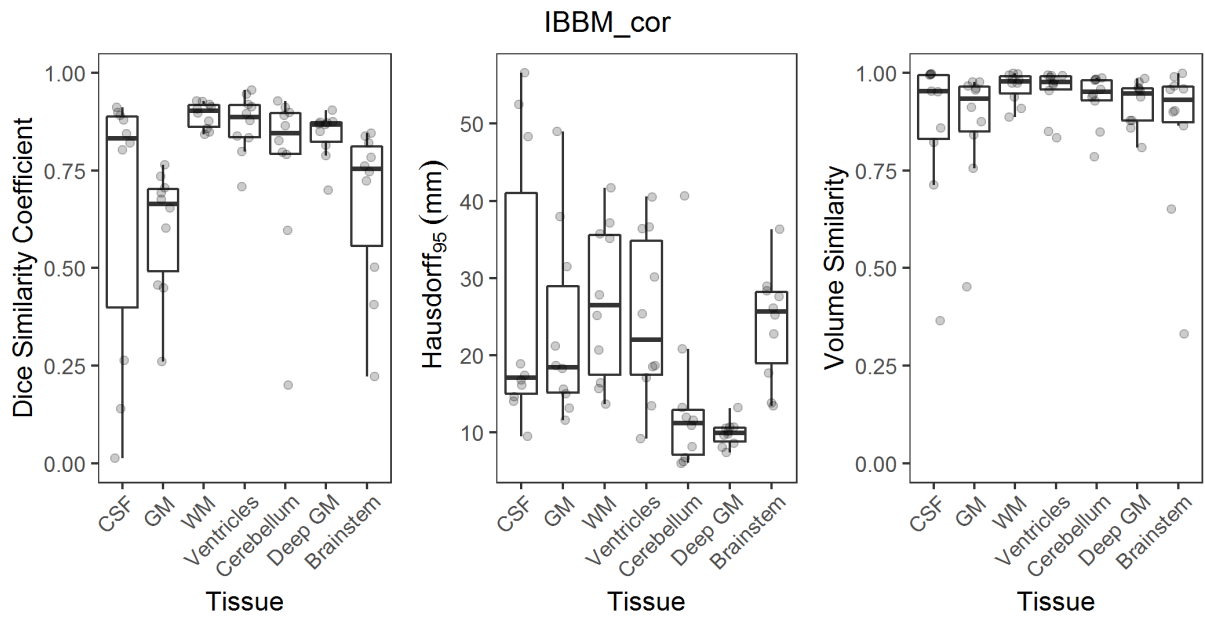

**IBBM Method - Combined**

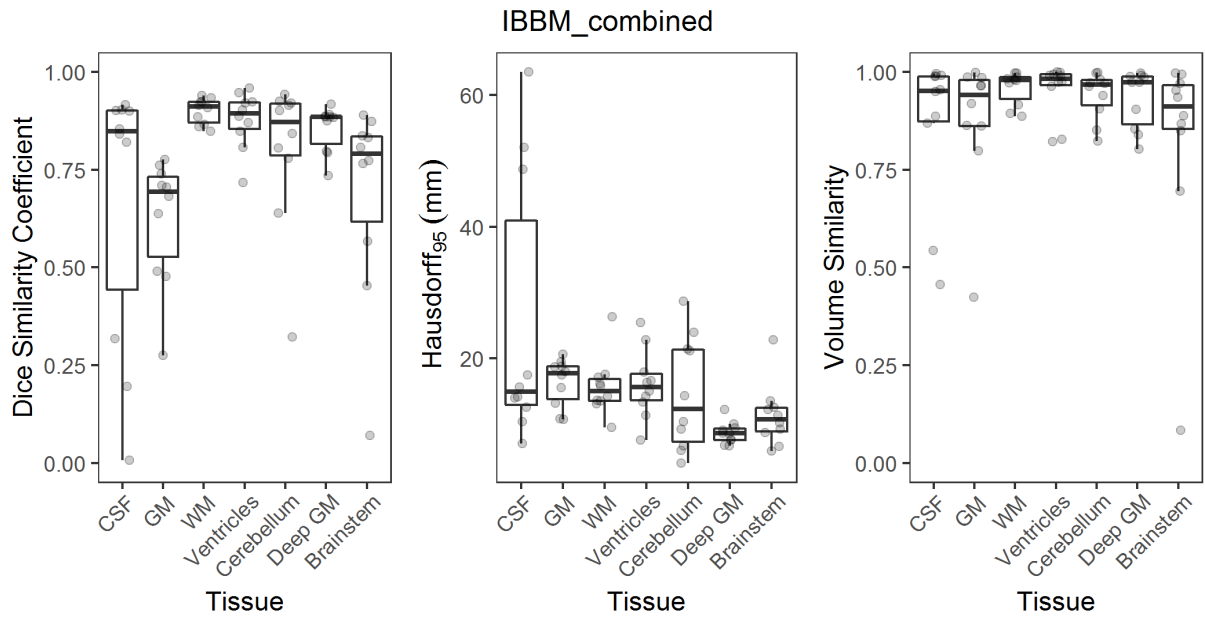

**Kispi Method - 2D axial orientation**

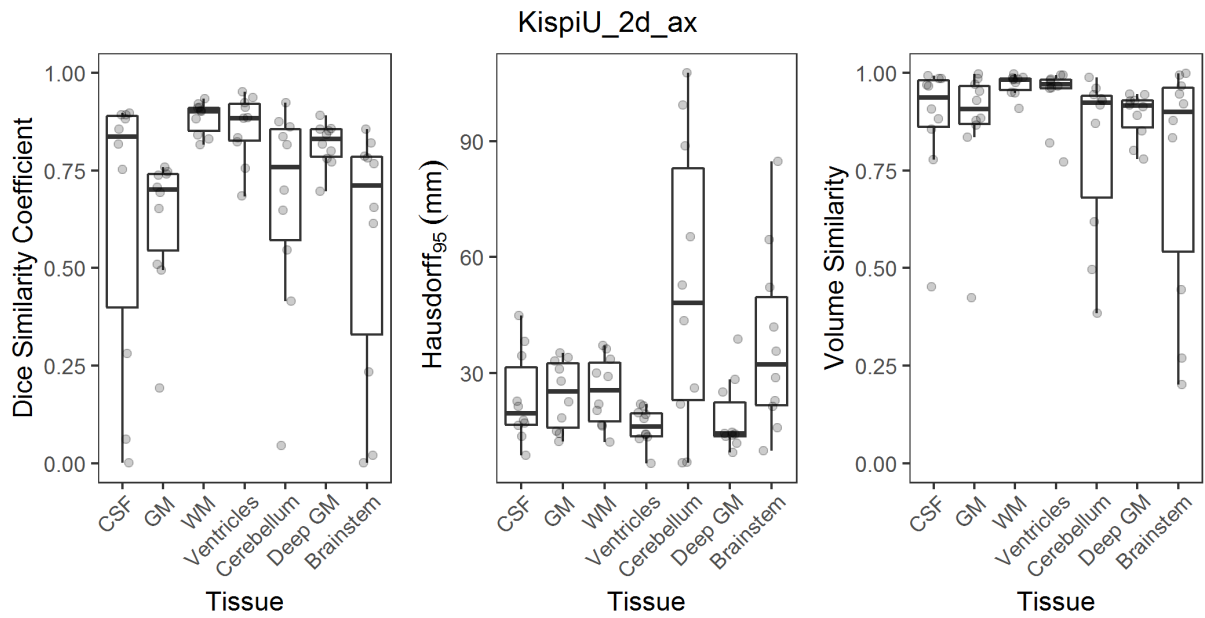

**Kispi Method - 2D sagittal orientation**

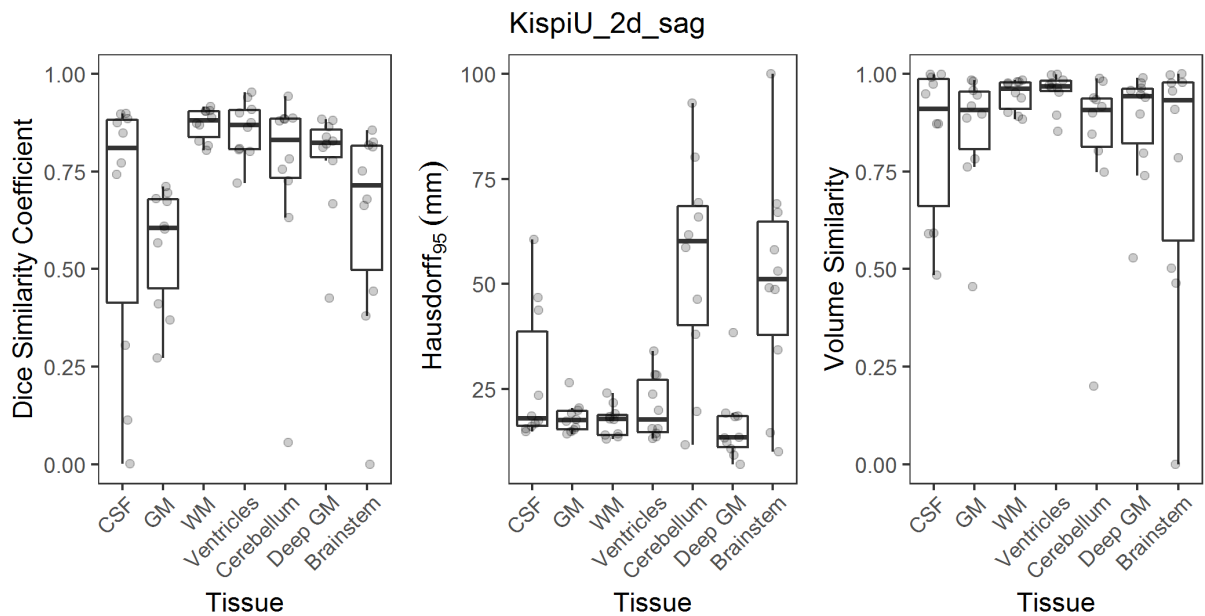

**Kispi Method - 2D coronal orientation**

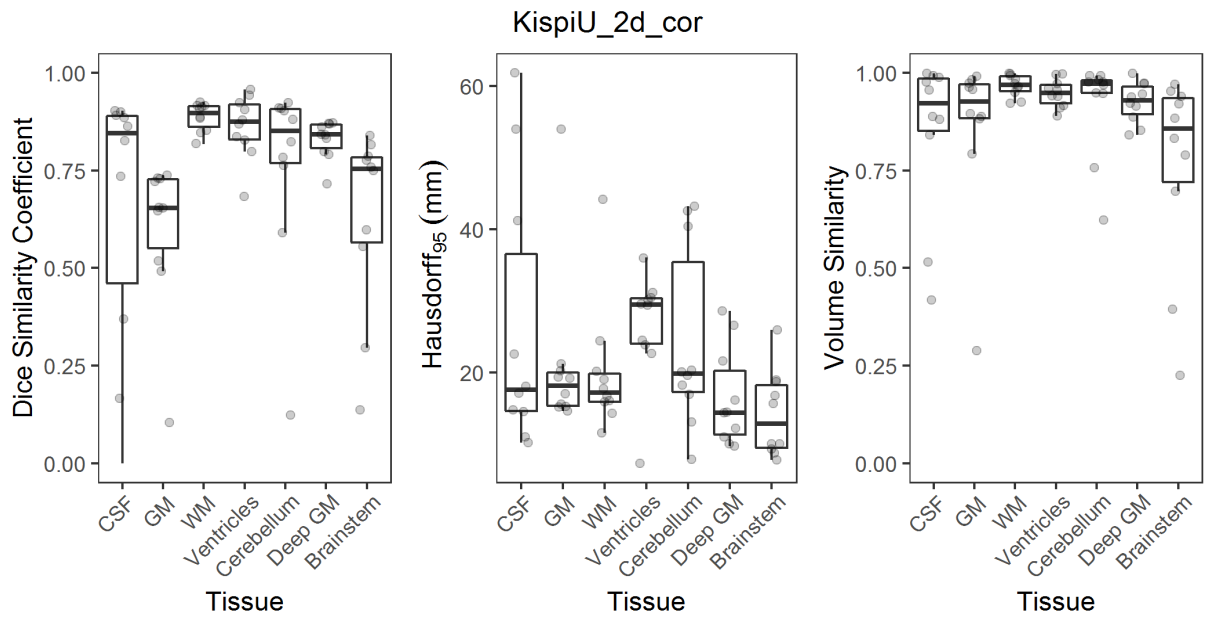

**Kispi Method - 3D orientation**

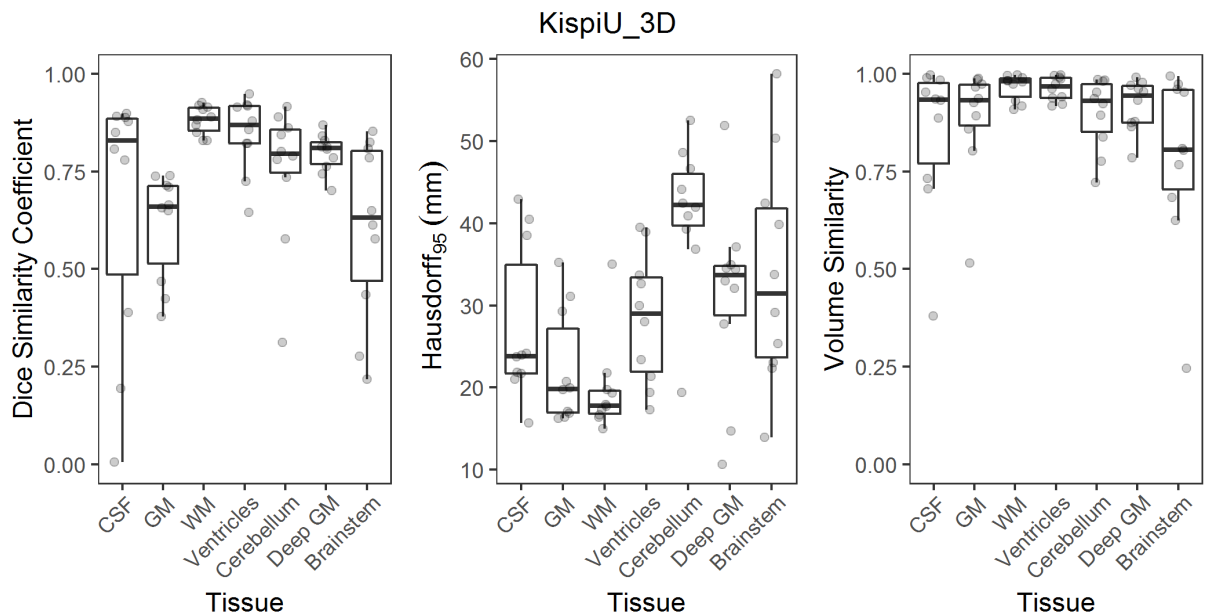

## Annotation Guidelines

Annotation Guidelines used for the creation of the label maps can be found in the database on Synapse:
(website: http://dx.doi.org/10.7303/syn23747212).